\documentclass[12pt]{iopart}
\usepackage{epsfig}
\usepackage{citesort}

\newcommand{\gtrsim}{\,\rlap{\lower3.7pt\hbox{$\mathchar\sim$}}
\raise1pt\hbox{$>$}\,}
\newcommand{\lesssim}{\,\rlap{\lower3.7pt\hbox{$\mathchar\sim$}}
\raise1pt\hbox{$<$}\,}

\begin{document}

\title{Structure formation with strongly interacting neutrinos - implications for the cosmological neutrino mass bound}
\author{Steen Hannestad \footnote{hannestad@fysik.sdu.dk}}
\address{Physics Department, University of Southern Denmark\\
Campusvej 55, DK-5230 Odense M, Denmark}

\date{{\today}}

\begin{abstract}
We investigate a model where neutrinos are strongly coupled to a new, light scalar field. In this model neutrinos annihilate as soon as they become non-relativistic in the early universe, and a non-zero neutrino mass has a marginal effect on the matter power spectrum. However, the angular power spectrum of the cosmic microwave background (CMB) is changed significantly because the strongly interacting fluid of neutrinos and scalars does not experience any anisotropic stress. Such models are strongly disfavoured by current observations. Interestingly, this leads to the conclusion that the relativistic energy density around the epoch of recombination must be in the form of very weakly interacting particles. This conclusion is independent of the specific interaction model.
\end{abstract}
\maketitle

\section{Introduction} 

The standard cosmological $\Lambda$CDM model with very light, weakly interacting neutrinos provides a very good fit to all current cosmological observations using only 6 parameters.
Adding a significant neutrino mass of order 1 eV worsens the fit and in fact allows for strong cosmological bounds on the sum of all active neutrino rest masses. Within the $\Lambda$CDM model the upper bound at present is around $\sum m_\nu \lesssim 0.5-1$ eV, depending on the specific priors and data used 
\cite{Spergel:2003cb,Hannestad:2003xv,Allen:2003pt,Barger:2003vs,%
Hannestad:2003ye,Crotty:2004gm,Hannestad:2004nb,sel04,Fogli:2004as,%
Hannestad:2004bu,Elgaroy:2003yh}.
This limit is almost an order of magnitude stronger than the current upper bound on the effective electron neutrino mass from beta decays.
Tritium decay measurements have been able to put an upper limit on the
mass of 2.3 eV (95\% conf.) \cite{kraus}, which translates into an upper bound on the sum of neutrino masses of roughly 7 eV.

Very interestingly there is also a claim of direct detection of
neutrinoless double beta decay in the Heidelberg-Moscow experiment
\cite{Klapdor-Kleingrothaus:2001ke,Klapdor-Kleingrothaus:2004wj},
corresponding to an effective neutrino mass in the $0.1-0.9$ eV
range. If this result is confirmed then it shows that neutrino
masses are almost degenerate and well within reach of cosmological
detection in the near future.

However, the cosmological neutrino mass bounds do depend on a number of assumptions, and are thus not on quite the same footing as the direct laboratory measurements. Many different possibilities have been discussed for alleviating the cosmological mass bound. The conclusion is that the mass bound is surprisingly robust. For instance, including cosmological defects has little effect \cite{Brandenberger:2004kc}. Another possibility, which has so far not been investigated, would be if the universe exits reheating at extremely low temperatures. In this case neutrinos would not fully thermalize before they decouple, and the end result would be that the neutrino number density is smaller than in the standard model \cite{Giudice:2000ex,Kawasaki:2000en,Adhya:2003tr,%
Hannestad:2004px,Gelmini:2004ah}.

Recently it was also proposed that neutrinos could be interacting strongly with a new, massless scalar field \cite{Beacom:2004yd}. In this model, neutrinos annihilate as soon the temperature drops below the neutrino rest mass and transfer all their entropy to the scalars.
The interesting feature of this model is that the neutrino rest mass never plays a significant role during structure formation so that cosmological neutrino mass bounds could disappear entirely. In the present paper we investigate this model in detail by solving the Boltzmann equation for a strongly interacting fluid of neutrinos and scalar. We find that, although the matter power spectra are indeed almost unaffected by the introduction of neutrino-scalar interactions, the spectrum of cosmic microwave background fluctuations is changed significantly.
In the next section we discuss the formalism needed to study structure formation with a neutrino-scalar fluid. Section 3 discusses a likelihood analysis based on present CMB and large scale structure data, and finally section 4 contains a discussion.

\section{General scenario}

The model proposed in Ref.~\cite{Beacom:2004yd} introduces a new scalar field which interacts with neutrinos via a simple scalar or pseudoscalar coupling,
\begin{equation}
{\cal L} = h_{ij} \bar{\nu}_i \nu_j \phi + g_{ij} \bar{\nu}_i \gamma_5 \nu_j \phi.
\end{equation}
Here we do not discuss the details of model building, but simply assume that such a coupling exists. In order to allow for almost complete annihilation of neutrinos at the present, the dimensionless coupling constant, $g_{ij}$, should be $g_{ij} \gtrsim 10^{-5}$ for all neutrino flavours. 

On the other hand, as discussed in \cite{Beacom:2004yd}, couplings much larger than this are strongly disfavoured by neutrinoless double beta decay and by supernova observations.
We therefore take $g \sim 10^{-5}$ to be characteristic of the model.
As was also calculated in Ref.~\cite{Beacom:2004yd}, the annihilation rate of neutrinos in the non-relativistic regime is roughly 
\begin{equation}
\Gamma = \frac{g^4}{64 \pi} \frac{T}{m_\nu^3} \left(\frac{m_\nu T}{2 \pi}\right)^{3/2} e^{-m_\nu/T}.
\end{equation}
This should be compared with the Hubble expansion rate which is roughly given by $H \simeq 4.9 \times 10^{-12} \, h \, T_{\rm eV}^{3/2} \,\, {\rm s}^{-1}$ around recombination.

\begin{figure}[htb]
\hspace*{1.5cm}\includegraphics[width=100mm]{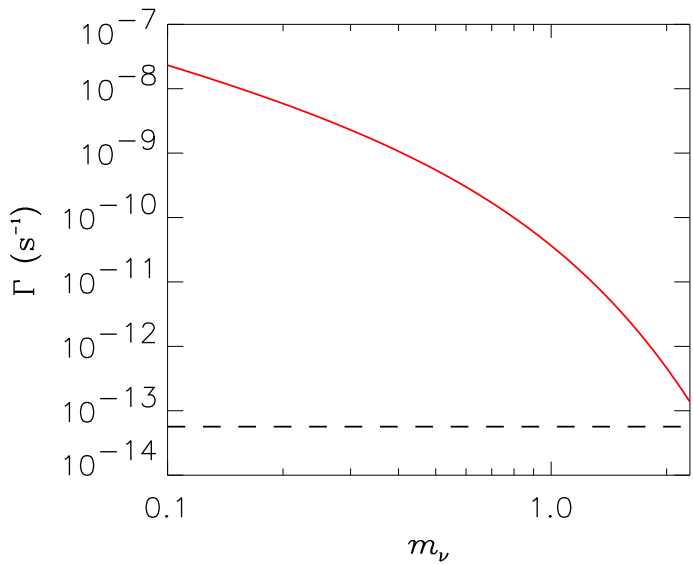}
\caption{The neutrino annihilation rate at $T = 0.3$ eV for different neutrino masses for $g=10^{-5}$. The horizontal dashed line is $H$ at $T = 0.3$ eV.}
\label{fig:gamma}
\end{figure}

As can be seen, even for neutrino masses which are close to the upper bound from $\beta$-decay measurements, the annihilation rate is much larger than the Hubble expansion rate until after recombination occurs at $T \simeq 0.3$ eV.
This means that in this model neutrinos and scalars can be treated as a perfect fluid with effectively zero mean free path for purposes of CMB and structure formation calculations. As will be discussed in the next section this has significant implications for CMB and structure formation.

\section{The Boltzmann equation}
\label{sec:GRboltz}

The evolution of any given particle species can be described via the Boltzmann equation. Our notation is identical to that of Ma and Bertschinger (MB) \cite{mb}.
The simplest choice is to use synchronous gauge because the numerical routine for calculating matter and CMB power spectra, CMBFAST \cite{CMBFAST}, is written in this gauge.
As the time variable we use conformal time, defined as $d \tau = dt/a(t)$, where $a(t)$ is the scale factor. Also, as the momentum variable we shall use the
comoving momentum $q_j \equiv a p_j$. We further parametrize $q_j$ as
$q_j = q n_j$, where $q$ is the magnitude of the comoving momentum and
$n_j$ is a unit 3-vector specifying direction.

The Boltzmann equation can generically be written as
\begin{equation}
L[f] = \frac{Df}{D\tau} = C[f],
\end{equation}
where $L[f]$ is the Liouville operator.
The collision operator on the right-hand side describes
any possible collisional interactions.

One can then write the distribution function as
\begin{equation}
f(x^i,q,n_j,\tau) = f_0(q) [1+\Psi(x^i,q,n_j,\tau)],
\end{equation}
where $f_0(q)$ is the unperturbed distribution function.

This applies to neutrinos as well as to the scalars, but with different $f_0$ and $\Psi$.

In synchronous gauge the Boltzmann equation can be written as an evolution equation for $\Psi$ in $k$-space \cite{mb}
\begin{equation}
\frac{1}{f_0} L[f] = \frac{\partial \Psi}{\partial \tau} + i \frac{q}{\epsilon}
\mu \Psi + \frac{d \ln f_0}{d \ln q} \left[\dot{\eta}-\frac{\dot{h}+6\dot{\eta}}
{2} \mu^2 \right] = \frac{1}{f_0} C[f],
\label{eq:boltzX}
\end{equation}
where $\mu \equiv n^j \hat{k}_j$ and $\epsilon = (q^2+a^2 m^2)^{1/2}$.
$h$ and $\eta$ are the metric perturbations, defined from the perturbed space-time
metric in synchronous gauge \cite{mb}
\begin{equation}
ds^2 = a^2(\tau) [-d\tau^2 + (\delta_{ij} + h_{ij})dx^i dx^j],
\end{equation}
\begin{equation}
h_{ij} = \int d^3 k e^{i \vec{k}\cdot\vec{x}}\left(\hat{k}_i \hat{k}_j h(\vec{k},\tau)
+(\hat{k}_i \hat{k}_j - \frac{1}{3} \delta_{ij}) 6 \eta (\vec{k},\tau) \right).
\end{equation}

\subsection{Collisionless Boltzmann equation}

At first we assume that $\frac{1}{f_0} C[f] = 0$. This assumption will be relaxed later.
The perturbation is then expanded as
\begin{equation}
\Psi = \sum_{l=0}^{\infty}(-i)^l(2l+1)\Psi_l P_l(\mu).
\end{equation}
One can then write the collisionless
Boltzmann equation as a moment hierarchy for the $\Psi_l$
by performing the angular integration of $L[f]$
\begin{eqnarray}
\dot\Psi_0 & = & -k \frac{q}{\epsilon} \Psi_1 + \frac{1}{6} \dot{h} \frac{d \ln f_0}
{d \ln q} \label{eq:psi0}\\
\dot\Psi_1 & = & k \frac{q}{3 \epsilon}(\Psi_0 - 2 \Psi_2) \label{eq:psi1}\\
\dot\Psi_2 & = & k \frac{q}{5 \epsilon}(2 \Psi_1 - 3 \Psi_3) - \left(\frac{1}{15}
\dot{h}+\frac{2}{5}\dot\eta\right)\frac{d \ln f_0}{d \ln q} \\
\dot\Psi_l & = & k \frac{q}{(2l+1)\epsilon}(l \Psi_{l-1} - (l+1)\Psi_{l+1})
\,\,\, , \,\,\, l \geq 3
\end{eqnarray}
It should be noted here that the first two hierarchy equations are directly
related to the energy-momentum conservation equation.
This can be seen in the following way. Let us define the density and
pressure perturbations of the dark matter fluid as \cite{mb}
\begin{eqnarray}
\delta & \equiv & \delta \rho/\rho \\
\theta & \equiv & i k_j \delta T^0_j/(\rho+P) \\
\sigma & \equiv & -(\hat{k}_i \hat{k}_j - \frac{1}{3} \delta_{ij})
(T^{ij}-\delta^{ij}T^k_k/3).
\end{eqnarray}
Then energy and momentum conservation implies that \cite{mb}
\begin{eqnarray}
\dot\delta & = & -(1+w)\left(\theta+\frac{\dot h}{2}\right)-
3 \frac{\dot a}{a} \left(\frac{\delta P}{\delta \rho} - w \right) \delta \label{eq:pert1} \\
\dot \theta & = & \frac{\dot a}{a} (1-3 w)\theta - \frac{\dot w}{1+w}
\theta + \frac{\delta P/ \delta \rho}{1+w} k^2 \delta - k^2 \sigma.
\label{eq:pert2}
\end{eqnarray}
By integrating Eq.~(\ref{eq:psi0}) over $q^2 \epsilon dq$,
Eq.~(\ref{eq:pert1}) is derived and by integrating Eq.~(\ref{eq:psi1})
equation over $q^3 dq$ one retrieves Eq.~(\ref{eq:pert2}).

\subsection{Collisional Boltzmann equation}

Starting from the general Boltzmann equation in synchronous gauge, Eq.~(\ref{eq:boltzX}), we now introduce
interactions by lifting the restriction that $\frac{1}{f_0} C[f] =
0$. Ideally, one should calculate the collision integrals in
detail for some explicit interaction. However, for very strongly interacting neutrinos and scalars it is in fact quite simple to understand how the fluid behaves.

The collision term in Eq.~(\ref{eq:psi0}) is
$\int d \Omega \frac{1}{f_0} C[f]$ and the one in
Eq.~(\ref{eq:psi1}) is $\int d \Omega \mu \frac{1}{f_0} C[f]$.
Integrating these two terms over momentum space one gets the
collision terms in Eqs.~(\ref{eq:pert1}-\ref{eq:pert2}) to be
\begin{equation}
\int C[f] d \Omega q^2 dq \epsilon
\end{equation}
and
\begin{equation}
\int C[f] d \Omega q^2 dq \mu q = k^i \int C[f] d \Omega q^2 dq
q_i
\end{equation}
respectively. 

Looking at particle species $i$, for processes of type $ii \to ii$, any integral of the form
\begin{equation}
\int C[f] d \Omega q^2 dq A,
\end{equation}
where $A \in (I,\epsilon,q_i)$ is automatically zero because $A$
is a collisional invariant.
Thus, both the above integrals are zero, and the right hand side
of the $l=0$ and 1 terms should be zero, reflecting that energy
and momentum is conserved in each interaction.

All the higher order terms can be estimated from the relaxation time approximation \cite{Hannestad:2000gt}, in which
\begin{equation}
\frac{1}{f_0}C[f] = -\frac{\Psi}{\tau}.
\label{eq:reltime}
\end{equation}
Here, $\tau$ is the mean time between collisions. Since the neutrino-scalar fluid is assumed to be very strongly interacting, the collision term dominates all other terms in the Boltzmann equation (since $\tau \ll H^{-1}$) unless $\Psi = 0$. 
This argument applies to neutrinos as well as scalars, so that for $l \geq 2$, $\Psi_{\nu,l} = \Psi_{\phi,l}=0$. In terms of the momentum-integrated equations, this means that $\sigma = 0$, so that the fluid experiences no anisotropic stress.

Since there are also creation and annihilation processes, $ii \to jj$,
there will be non-zero terms for $l=0$ and 1, of the form \cite{Kaplinghat:1999xy}
\begin{equation}
\left. \dot\Psi_{l,i}\right|_{\rm coll} = \frac{1}{\tau}(\Psi_{l,j}-\Psi_{l,i}),
\end{equation}
equivalent to the relaxation time approximation for $ii \to ii$, Eq.~(\ref{eq:reltime}).
Again, since $\tau \ll H^{-1}$ this means that these terms dominate unless $\Psi_{l,j} = \Psi_{l,i}$ for $l=0,1$.

Altogether, this means that the combined neutrino-scalar fluid can be treated as a single fluid with a modified effective equation of state,
\begin{equation}
w_{\rm eff} = \frac{P_\nu + P_\phi}{\rho_\nu + \rho_\phi}
\end{equation}
The perturbation equations to be solved are Eqs.~(\ref{eq:pert1}-\ref{eq:pert2}) with $w = w_{\rm eff}$.
This is exactly equivalent to the fluid of electrons, protons and neutral hydrogen around the epoch of recombination which can also be considered a single fluid because the interaction time scale is much shorter than the expansion timescale \cite{Hannestad:2000fy}.

The effective equation of state is then straightforward to calculate. The temperature of the neutrino-scalar fluid, $T_{\nu \phi}$, is found by solving Friedmann equation and the equation of energy conservation
\begin{eqnarray}
H^2 & = & \frac{8 \pi G (\rho_\nu+\rho_\phi+\rho_\gamma+\rho_m+\rho_\Lambda)}{3} \\
\frac{dT_{\nu \phi}}{dt} & = & -H \frac{4 \rho_\phi + 3(\rho_\nu + P_\nu)}{d(\rho_\nu+\rho_\phi)/dT_{\nu \phi}},
\end{eqnarray}
with the energy density and pressure of each species given by
\begin{eqnarray}
\rho_\nu & = & \frac{3}{\pi^2}\int_0^\infty \frac{p^2 E dp}{e^{E/T_{\nu \phi}}+1} \\
P_\nu & = & \frac{3}{\pi^2}\int_0^\infty \frac{p^4/(3E) dp}{e^{E/T_{\nu \phi}}+1} \\
\rho_\phi & = & \frac{\pi^2}{30} T_{\nu \phi}^4 \\
P_\phi & = & \rho_\phi/3
\end{eqnarray}

\begin{figure}[htb]
\hspace*{1.5cm}\includegraphics[width=100mm]{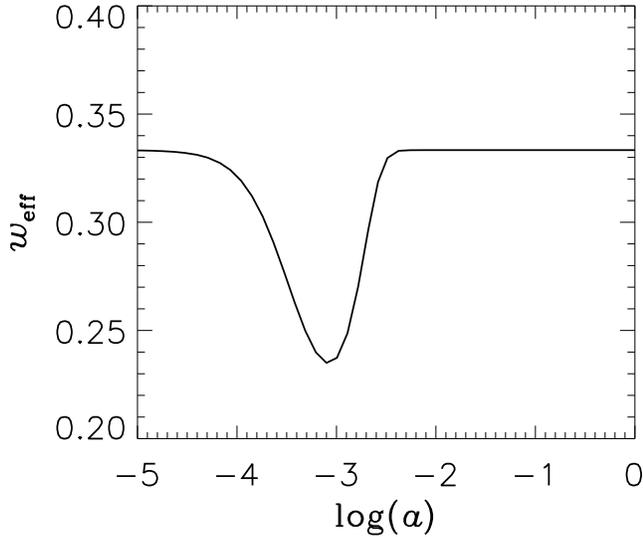}
\caption{The effective equation of state for the combined neutrino-scalar fluid for a neutrino mass of 1 eV.}
\label{fig:w}
\end{figure}

In Fig.~\ref{fig:w} we show the effective equation of state of the combined neutrino-scalar fluid for a neutrino mass of 1 eV.
At $T \gg m_\nu$ and $T \ll m_\nu$ the fluid is extremely relativistic so that $w = 1/3$. During the annihilation epoch when the rest mass of the massive neutrinos plays a role, the pressure is lower.

\begin{figure}[htb]
\hspace*{1.5cm}\includegraphics[width=100mm]{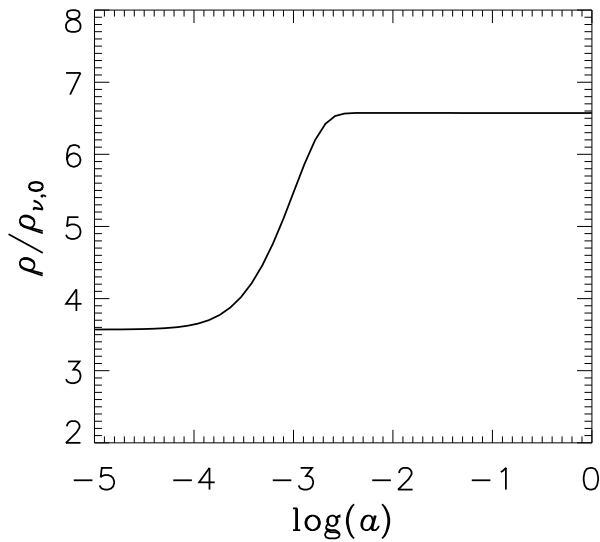}
\caption{The total energy density in the combined neutrino-scalar fluid for a neutrino mass of 1 eV.}
\label{fig:rho}
\end{figure}

In Fig.~\ref{fig:rho} we show the energy density of neutrinos and scalars for the same model, normalized to that of a massless standard model neutrino. At high temperatures, $\rho/\rho_{\nu,0} =  3 + \frac{4}{7} \simeq 3.57$, whereas after neutrino annihilation it is $\frac{1}{2} \times \frac{8}{7} \times \left(25/4\right)^{4/3} \simeq 6.58$.

\section{Matter power spectra}

We have calculated matter power spectra for models with a fluid of strongly interacting neutrinos and scalars. 

\begin{figure}[htb]
\hspace*{1.5cm}\includegraphics[width=100mm]{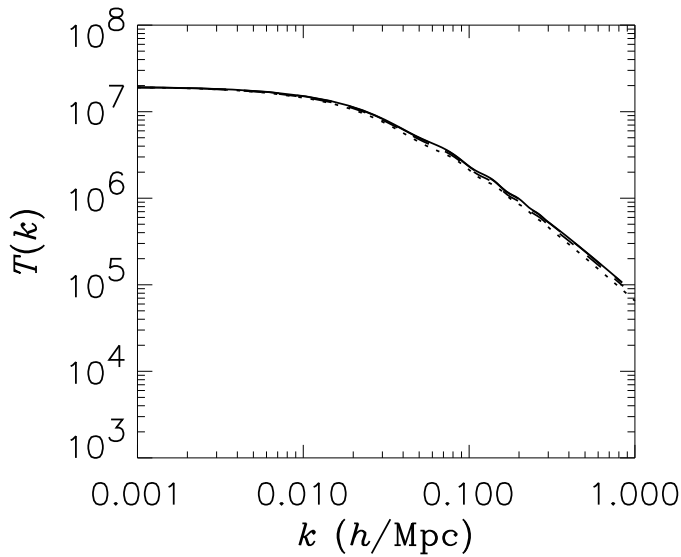}
\caption{Transfer functions for $\Lambda$CDM models with neutrino masses of 0.01 eV (solid), 0.1 eV (long-dashed), 1 eV (dashed), and 10 eV (dotted), respectively.}
\label{fig:fig3}
\end{figure}

Instead of plotting the power spectrum, $P(k)$, we plot the transfer function $T(k)$ which is defined according to $P(k) = T^2(k) P_0(k)$.
$P_0(k)$ is the primordial power spectrum, which is independent of neutrino physics.
We show these transfer function in Fig.~\ref{fig:fig3} for a standard $\Lambda$CDM model with parameters
$\Omega = \Omega_m + \Omega_\Lambda = 1$, $\Omega_m = 0.3$, $\Omega_b = 0.05$, $H_0 = 70 \, {\rm km} \, {\rm s}^{-1} \, {\rm Mpc}^{-1}$, $n_s = 1$, and $\tau = 0$. 

The stated mass is for a single mass eigenstate, and all mass eigenstates are assumed degenerate so that $\sum m_\nu = 3 m_\nu$. Note that for $m_\nu = 0.01$ eV this is unphysical since the mass difference $\delta m_{23}$ is known from oscillation experiments to be of order 0.05 eV \cite{Maltoni:2004ei,Maltoni:2003da,Aliani:2003ns,deHolanda:2003nj}. However, this is of no qualitative importance for the analysis.

From this figure it is clear that the matter power spectrum is not changed significantly, even for very large neutrino masses. This is in accordance with the finding in Ref.~\cite{Beacom:2004yd}. However, in Ref.~\cite{Beacom:2004yd} only the energy density in the neutrino-scalar fluid was accounted for. The strong neutrino-scalar interaction means that the fluid does not free-stream, but rather undergoes acoustic oscillations. 

\begin{figure}[htb]
\hspace*{1.5cm}\includegraphics[width=100mm]{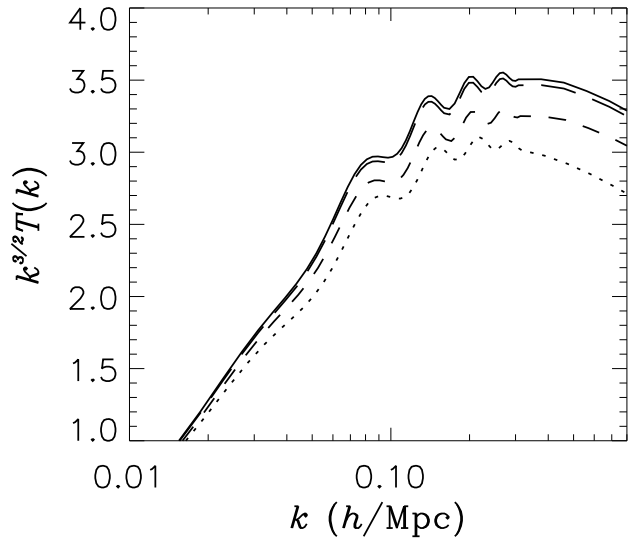}
\caption{Transfer functions for $\Lambda$CDM models with neutrino masses of 0.01 eV (solid), 0.1 eV (long-dashed), 1 eV (dashed), and 10 eV (dotted), respectively. The normalization is arbitrary.}
\label{fig:fig7}
\end{figure}

In Fig.~\ref{fig:fig7} we show the difference between the different power spectra on small scales. The difference on small scales between $m_\nu = 0$ and $m_\nu \to \infty$ is roughly $T(m_\nu \to \infty)/T(m_\nu = 0) \simeq 0.8$, leading to the conclusion that the difference in the amplitudes of the matter power spectra is $T^2(m_\nu \to \infty)/T^2(m_\nu = 0) \sim 0.6$. This is in almost perfect agreement with Ref.~\cite{Beacom:2004yd}, where the ratio was found to be $0.65$.

\section{CMB power spectra}

Contrary to the matter power spectra there is a significant effect on the CMB, depending on the neutrino mass. If the neutrino mass is large, neutrinos annihilate before matter-radiation equality, delaying it and thereby increasing the early integrated Sachs-Wolfe effect.

\begin{figure}
\hspace*{1.5cm}\includegraphics[width=100mm]{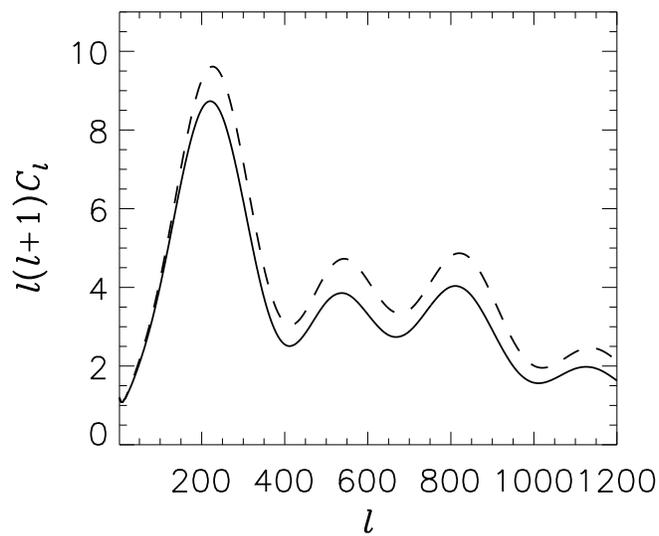}
\caption{CMB TT spectrum for a standard $\Lambda$CDM model with free-streaming (solid line) and strongly interacting (dashed line) neutrinos.}
\label{fig:fig1}
\end{figure}

However, there is also a significant effect from the fact that there is no stress term damping the neutrino-scalar oscillations.
This effectively increases the fluctuation level on scales which are sub-horizon prior to recombination. In Fig.~\ref{fig:fig1} this effect has been illustrated by simply setting all $l \geq 2$ modes equal to zero for standard, massless neutrinos. Physically this corresponds to having neutrinos undergo acoustic oscillations instead of free-streaming.

As can be seen from the figure, the effect is very similar to introducing a step-like feature in the power spectrum which would increase power on all scales smaller than $l_{\rm break}$, which should then be of order the horizon scale at decoupling.

\begin{figure}
\hspace*{1.5cm}\includegraphics[width=100mm]{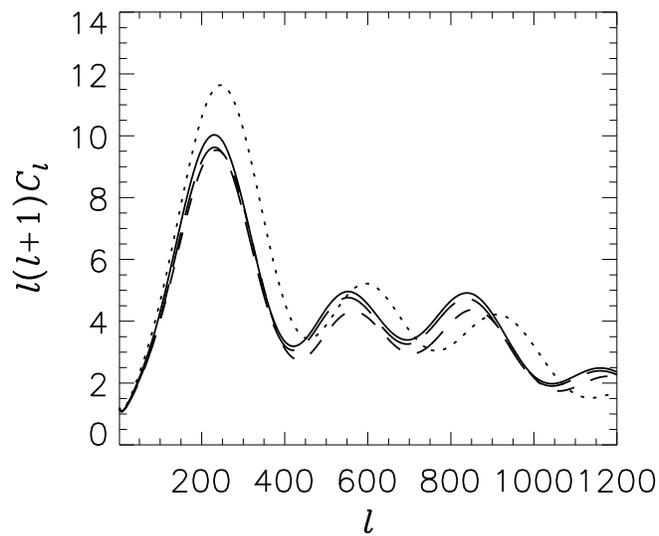}
\caption{CMB TT power spectra for $\Lambda$CDM models with neutrino masses of 0.01 eV (solid), 0.1 eV (long-dashed), 1 eV (dashed), and 10 eV (dotted), respectively.}
\label{fig:fig4}
\end{figure}

In Fig.~\ref{fig:fig4} we show CMB spectra for same models as in Fig.~\ref{fig:fig3}. As can be seen there are several different effects interplaying. For the model with $m_\nu = 0.01$ eV, recombination occurs long before annihilation, but for the models with $m_\nu = 0.1$ and 1 eV annihilation happens around the same epoch as recombination. Since the effective $w$ is lower at this point there is a slight decrease in the early ISW effect compared to the $m_\nu = 0.01$ eV model. Finally, in the model with $m_\nu = 10$ eV annihilation occurs long before recombination so that the early ISW effect is strongly enhanced because of the additional radiation present during recombination.

The fact that the CMB spectrum is changed substantially makes it difficult to predict whether these models are in fact capable of fitting current data as well as the standard $\Lambda$CDM model. Since the CMB fluctuations are enhanced on sub-horizon scales, one prediction is that models with higher $\Omega_m$ and/or $H_0$ will produce better fits since such models produce less power around the first acoustic peak.
In the next section we present a likelihood analysis of present CMB and large scale structure data for models with a strongly interacting neutrino-scalar fluid.

\section{Likelihood analysis}

In order to test how well models with a strongly interacting neutrino-scalar fluid fit data, we have performed a likelihood analysis of CMB and LSS data.

As our framework we choose the
minimum standard model with 6
parameters: $\Omega_m$, the matter density, $\Omega_b$, the baryon density, $H_0$, the Hubble
parameter, and $\tau$, the optical depth to reionization. The normalization of both CMB and LSS spectra are taken to be free and unrelated parameters. The only additional parameter is then the neutrino mass, $m_\nu$. We assume that there are three active neutrino mass eigenstates with degenerate masses so that $\sum m_\nu = 3 m_\nu$.

\begin{table}[ht]
\caption{\label{tab:priors} Priors on cosmological parameters used in
the likelihood analysis.}
\begin{indented}
\item[]
\begin{tabular}{@{}lllll}
\br
&\multicolumn{2}{l}{CMB only}&\multicolumn{2}{l}{CMB+LSS+HST} \cr
\br
Parameter &Prior&Distribution&Prior&Distribution\cr
\mr
$\Omega=\Omega_m+\Omega_X$&1&Fixed&1&Fixed\\
$h$ & 0.4-1.1 & Top hat & $0.72 \pm 0.08$&Gaussian \cite{freedman}\\
$\Omega_b h^2$ & 0.014--0.040&Top hat & 0.014--0.040&Top hat\\
$n_s$ & 0.6--1.4& Top hat & 0.6--1.4& Top hat\\
$\tau$ & 0--1 &Top hat & 0--1 &Top hat\\
$Q$ & --- &Free & --- &Free \\
$b$ & --- & --- & --- &Free\\
\br
\end{tabular}
\end{indented}
\end{table}

Likelihoods are calculated from $\chi^2$ so that for 1 parameter estimates, 68\% confidence regions are determined by $\Delta \chi^2 = \chi^2 - \chi_0^2 = 1$, and 95\% region by $\Delta \chi^2 = 4$. $\chi_0^2$ is $\chi^2$ for the best fit model found.

\subsection{Large Scale Structure (LSS).}

At present there are two large galaxy surveys of comparable size, the
Sloan Digital Sky Survey (SDSS) \cite{Tegmark:2003uf,Tegmark:2003ud}
and the 2dFGRS (2~degree Field Galaxy Redshift Survey) \cite{2dFGRS}.
Once the SDSS is completed in 2005 it will be significantly larger and
more accurate than the 2dFGRS. In the present analysis we use data from SDSS, but the results would be almost identical had we used 2dF data instead. In the data analysis we use only data points on scales larger than $k = 0.15 h$/Mpc in order to avoid problems with non-linearity.

\subsection{Cosmic Microwave Background.}

The CMB temperature fluctuations are conveniently described in terms of
the spherical harmonics power spectrum $C_l^{TT} \equiv \langle
|a_{lm}|^2 \rangle$, where $\frac{\Delta T}{T} (\theta,\phi) =
\sum_{lm} a_{lm}Y_{lm}(\theta,\phi)$.  Since Thomson scattering
polarizes light, there are also power spectra coming from the
polarization. The polarization can be divided into a curl-free $(E)$
and a curl $(B)$ component, yielding four independent power spectra:
$C_l^{TT}$, $C_l^{EE}$, $C_l^{BB}$, and the $T$-$E$ cross-correlation
$C_l^{TE}$.

The WMAP experiment has reported data only on $C_l^{TT}$ and $C_l^{TE}$
as described in
Refs.~\cite{Bennett:2003bz,Spergel:2003cb,%
Verde:2003ey,Kogut:2003et,Hinshaw:2003ex}.  We have performed our
likelihood analysis using the prescription given by the WMAP
collaboration~\cite{Spergel:2003cb,%
Verde:2003ey,Kogut:2003et,Hinshaw:2003ex} which includes the
correlation between different $C_l$'s. Foreground contamination has
already been subtracted from their published data.

\subsection{Results}

We consider two cases in the likelihood analysis: (a) Only WMAP temperature and polarization data, (b) WMAP data and SDSS data, combined with data on the Hubble parameter from the HST key project \cite{freedman}, $h = H_0/(100 \, {\rm km} \, {\rm s}^{-1} \, {\rm Mpc}^{-1}) = 0.72 \pm 0.08$.
The priors we use are given in Table~\ref{tab:priors}.

\begin{figure}
\hspace*{1.5cm}\includegraphics[width=100mm]{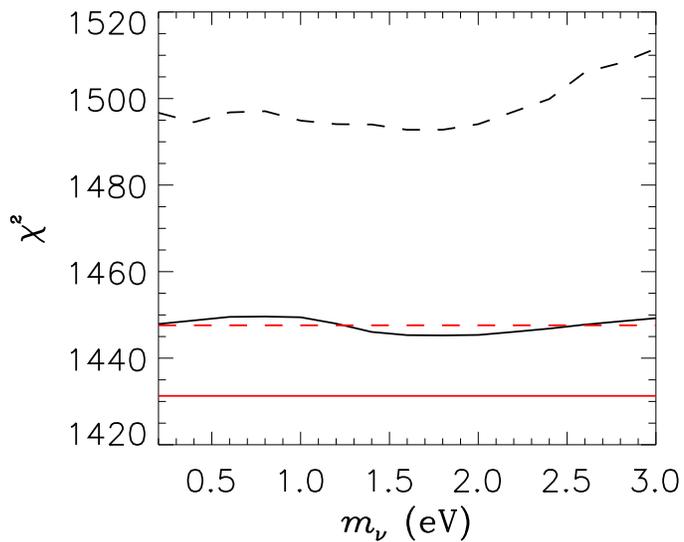}
\caption{$\chi^2$ for WMAP TT and TE data only (full line), and WMAP+SDSS+HST data (dashed line). The horizontal lines show best fit $\chi^2$ for $\Lambda$CDM models with massless neutrinos.}
\label{fig:chi2}
\end{figure}

In Fig.~\ref{fig:chi2} we show $\chi^2$ as a function of $m_\nu$, obtained from minimizing over all the other cosmological parameters.
As can be seen from the figure the fit to data is quite poor in both cases when compared to the standard $\Lambda$CDM model with non-interacting neutrinos, independent of neutrino mass.
Of course it might be possible to obtain a better fit by introducing additional parameters to the analysis, but this would be at the price of making the model more contrived.

\begin{figure}
\hspace*{1.5cm}\includegraphics[width=80mm]{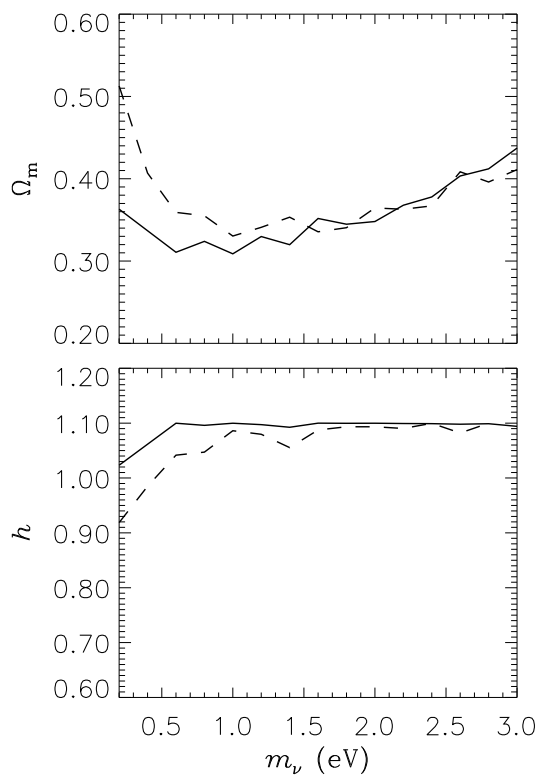}
\vspace*{1cm}
\caption{Values of $\Omega_m$ and $h$ for the best fit models, as a function of $m_\nu$, for WMAP TT and TE data only (full line), and WMAP+SDSS+HST data (dashed line).}
\label{fig:other}
\end{figure}

In Fig.~\ref{fig:other} we show best fit values of $\Omega_m$ and $h$ as functions of neutrino mass. As can be seen the best fit value of the Hubble parameter is highly non-standard for all values of the neutrino mass. In order to compensate for the increase in power due to the lack of free-streaming, the Hubble parameter has to be increased significantly in order to provide a decent fit. This in turn is in strong disagreement with measurements of the Hubble parameter from the HST key project \cite{freedman}

\section{Discussion}

We have calculated in detail how structure formation proceeds in models with a strongly interacting plasma of neutrinos and massless scalars.
These models differ from models with standard non-interacting neutrinos because neutrinos annihilate when the temperature drops below the rest mass. Furthermore, in these models the fluid of neutrinos and scalars undergoes acoustic oscillations instead of free-streaming.

The effect is that the matter power spectrum is only minimally altered by a non-zero neutrino rest mass, but that the CMB power spectrum changes significantly, both because of the additional relativistic energy density and the strong coupling of the neutrino-scalar fluid.

For standard, non-interacting neutrinos the neutrino free-streaming sets an upper limit on the neutrino rest mass when CMB and LSS data is combined. While neutrinos in the strongly interacting models do not suffer free-streaming and therefore produce matter power spectra which are in accordance with standard $\Lambda$CDM, it is impossible to simultaneously fit CMB and LSS data. Models with strongly interacting neutrinos and scalars are therefore strongly disfavoured by observations. Interestingly this also means that although present cosmological observations cannot distinguish between very light neutrinos and other types of relativistic energy density, it is possible to say that the relativistic energy density must be in the form of very weakly interacting particles which free-stream instead of oscillating acoustically (see also \cite{bs}). This is yet another confirmation of the standard picture of $\Lambda$CDM cosmology.

\section*{Acknowledgments} 

Use of the publicly available CMBFAST
package~\cite{CMBFAST} and of computing resources at DCSC (Danish
Center for Scientific Computing) are acknowledged. 

\vspace*{2cm}

\section*{References} 

\end{document}